\documentstyle[preprint,aps]{revtex}

\begin{document}  
\bibliographystyle{apsrev}
  
\title{ Tuning the superfluid - Mott insulator
transition 
in a resonantly strongly driven Bose - Einstein condensate
in an
optical lattice.}  
\author{G.M.Genkin$^*$.}
\address{  Physics Department
and Center for Polymer Studies,
Boston  University ,
              Boston, MA 02215.}

\maketitle                               
\begin{abstract}

   We have studied the superfluid - Mott insulator
    transition in a BEC
    confined
     in an optical lattice and driven by an additional external
    electromagnetic field, which
    excites another internal
    state. It is shown that due to resonant strong driving
    the critical values of the transition parameter  depend on 
    external field parameters (the Rabi frequency and the detuning from
    resonance)
       and can be tuned to a given value. It is also shown that
 for driven system there are  two optical potential
 depths at a phase transition, which are always more than without driving.

 {*} Electronic address: ggenkin@argento.bu.edu.
\end{abstract}  

\maketitle


   The experimental observation of Bose-Einstein condensation
  in a dilute gas of ultracold trapped atoms [ 1-3 ] has generated much interest
  in 
  manipulating such coherent matter by 
  external electromagnetic fields. 
 Recently, atoms have been confined in optical potentials created by
 standing light waves [ 4, 5 ], with the wavelength of the optical potential
 much smaller than the dimensions of the trap. Considerable work has been done,
 in the last decade, to determine the ground state phase diagram of correlated
 bosons on such a lattice. The bosons can undergo a transition from 
 superfluid to  insulator. Strong competition between potential and
 kinetic energies begets a quantum phase transition. The on - site repulsion can
 produce a Mott insulating phase with a quantum phase transition to a
 superfluid as the interaction strength weakens. A seminal experiment by
 Greiner and collaborators [5] demonstrated a quantum phase transition
 in a Bose - Einstein condensate ( BEC )
  from
 a superfluid ( SF ) state into a Mott insulator ( MI ) state, by
 varying the lattice laser intensity, as proposed theoretically in [ 6 ].
The quantum phase transitions have attracted much interest in recent years.
These transitions are accomplished by changing not the temperature, but some
parameter in the Hamiltonian of the system; for a BEC in an optical lattice
this parameter is the lattice laser intensity, for solid state systems, for
example, the magnetic field in a quantum - Hall system, or the charging
energy in Josephson - junction arrays.

   In this paper, we propose a manipulation of the quantum phase transition of
  a BEC in an optical lattice by an additional external electromagnetic field.
  This basic idea of my paper is using the external field to drive transitions
  between internal states for atoms in an optical lattice. This driving
  alter the balance between self - interactons of the atoms in lattice site
  and hopping of the atoms between the sites, and thereby modify the
  superfluid - Mott insulator phase transition.
  
   An applied resonant field excites another internal state of the
    BEC, which is
   why the populations of two coupled states are varied. Without
   driving a quantum phase transition in an one - component BEC is a 
   single - channel problem, but with resonant driving it
   becomes a two - channel problem, in which every channel has competition
   between
     kinetic and potential energies.
    The populations of the coupled states 
   are  
   varied. As a result, the effective on - site interaction and the effective
   tunneling in these two channels, which depend on the population, are also
   changed. However, due to this varying population  the variations of the
   on - site interaction and the tunneling are different; therefore, the
   balance between  kinetic and potential energies, which determines a
    phase transition, is altered. As a result, we have a
   modification
   of the quantum phase transition by the
   applied resonant field, which is determined by the two channels.
   It is shown that
    for each
   channel, the critical value of the ratio of the interaction strength and
   the tunneling
   depends on the population, which for the 
    strong driving is the Rabi oscillations.
   Therefore,   
    in the regime of the
   time - dependent Rabi oscillations 
   the MI - SF and the reverse SF - MI transitions
    occur during a period of the Rabi frequency.  
    The frequency of this oscillating superfluidity is determined by the
    driving strength. 
      In the steady state driving regime 
    the critical value of the ratio of the parameters is always increased,
    and two optical potential depths at the phase transition for the
    driven system are always
    higher than without driving.
    Therefore, this effect opens the possibility 
   creating a superfluid state from a initially Mott
   insulator state.
      The critical values of the driven system depend
    on the parameters of the external
          field ( the Rabi frequency and
    the detuning from resonance );
    in particular, the quantum phase transition can be induced by varying only
    the detuning with the fixed field amplitude.
     We can tune the critical values
    of the quantum phase transition to a given value.
     Our proposal of manipulating a quantum phase transition can be applied to
     solid state systems. For example, in a quantum - Hall system the
     resonant driving may be a novel way of modification a quantum phase transition
     besides using the magnetic field which controls the transition between
     different quantized Hall phases.   
    
    We consider a two - component
    Bose - Einstein condensate in an optical lattice in the presence of an 
    additional
    external
    electromagnetic field which couples the two components (the different internal
    states).
             The energy eigenstates are Bloch states, and an appropriate
     superposition of Bloch states yields a set of Wannier functions which are
     localized on the individual lattice sites. An external field couples the two states $ a $ and $ b $,
    and an atom is always in the superposition of these states. 
   The system is driven by a two - photon Raman pulse, the strength of which
   we characterize by the Rabi frequency $ \Omega_R $.
      We will consider
     as the two distinct
     internal states $ a $ and $ b $ coupled by the external
     time - dependent field two different vibrational states in the
     Wannier basis, where the initial state
     $ a $ is the lowest vibrational state of the harmonic oscillator with
     $ n = 0 $ and the state $ b $ is the excited vibrational state with
     $ n \neq 0 $. 
      Depending on the relative 
     strength of the atom - atom interaction, the system may lie in the Rabi
     regime or in the Josephson    
  regime. The equations of motion of a strongly ( the Rabi regime ) driven
  [ 7, 8 ] two - component BEC resemble the Bloch equations describing a
  driven two - level system if the 
  external driving field $ \hbar \Omega_R $ is much larger than the
  difference in the two - body interaction ( the " charging" [ 9 ] energy
  $ E_C $ which is proportional to $ U_a + U_b - 2 U_{ab} $ ).For the two - body interaction, we use a contact potential,
    $ V_l({\bf x} -{\bf x}^{\prime}) = 
    U_l \delta( {\bf x} - {\bf x}^{\prime}) $, here \{l = {a, b}\};     
    the coupling constants, $ U_l $, are expressed in terms of the $ s $ -wave   
    scattering lengths by $ U_l = \frac{ 4 \pi \hbar^2 a_l }{ m} $, where
    $ a_a $ and $ a_b $ are the scattering lengths for collisions between two
    atoms in states $ a $ and $ b $, respectively, while $ a_{ab} $ is the
    scattering length for collisions between atoms in different internal states.   
      The driving field is a spatially uniform field of the two - photon pulse of 
  Raman type with the wavelengths much larger than the dimensions of a BEC.
    The external
     resonant field excites the higher internal state, while, the
   energies involved in the one channel dynamics   
    to be small compared to the excitation energies to the higher internal
    state.
    In this case of the strong driving we can solve the problem
     in two steps. The first is taking account only of the strong external
     field $ \Omega_R $; in the result of this step we have the Bloch
   equations solution  $\varphi_l$
   where the functions $\varphi_l$ are determined by the parameters of the
   external field ( the Rabi frequency $ \Omega_R $ and the detuning
   $ \delta $ from resonance ). This Rabi
 solution of the Bloch equations obey the condition
   $ |\varphi_a|^2 + |\varphi_b|^2 = 1 $.
   And, the wave function of a two - level system is
   $$
   \Psi = \varphi_a \Psi_a + \varphi_b \Psi_b.   \eqno (1)
   $$
     In this two - state Hilbert space the system is described by the
     effective Hamiltonian $ H_{eff} $.
      And, a second step is taking account of the
   remaining terms of the Hamiltonian. As a result,    
   using Eq.(1)
   we can write the effective Hamiltonian
     in the second - quantization notations for the boson field operators
     $$
    H_{eff} = \int d^{3}x [ \Psi_{a}^{\dagger}({\bf x},t) 
    ( - \frac{ \hbar^2 \nabla^2}{ 2 m} +
     V_0({\bf x}) + \frac{ \hbar \delta }{ 2} ) \Psi_a({\bf x},t) \varphi_a^2 +
    \Psi_{b}^{\dagger}({\bf x},t) ( - \frac{ \hbar^2 \nabla^2}{ 2 m} +
    V_0({\bf x}) - \frac{ \hbar \delta }{ 2 } ) \Psi_b({\bf x},t) \varphi_b^2+    
    $$    
    $$
      \int d^{3}x d^{3}x^{\prime} [ \Psi_{a}^{\dagger}({\bf x},t)
   \Psi_{a}^{\dagger}({\bf x}^{\prime},t) V_a({\bf x} - {\bf x}^{\prime})
   \Psi_a({\bf x}^{\prime},t) \Psi_a({\bf x},t) \varphi_a^4 +
    $$
    $$
  \Psi_{b}^{\dagger}({\bf x},t) \Psi_{b}^{\dagger}({\bf x}^{\prime},t)
   V_b({\bf x} -
    {\bf x}^{\prime}) \Psi_b({\bf x}^{\prime},t) \Psi_b({\bf x},t) \varphi_b^4 +
    2 \Psi_{a}^{\dagger}({\bf x},t) \Psi_{b}^{\dagger}({\bf x}^{\prime},t)
   V_{ab}({\bf x} - {\bf x}^{\prime}) \Psi_b({\bf x}^{\prime},t)
   \Psi_a({\bf x},t) \varphi_a^2 \varphi_b^2 ],                \eqno (2)
    $$
   where $ V_0({\bf x}) $ is the optical lattice potential,  
 usually, the optical lattice potential has the form
 $ V_0({\bf x}) =\sum_{j=1}^{3} V_{jo} \sin^2 k x_j $     
     with
   wave vectors $ k = \frac{ 2 \pi }{ \lambda} $ and $ \lambda $  the wavelength of 
   the laser beams.    
         The atomic field operators   
    have been written in the rotating    
     wave approximation and, consequently, in Eq.(2) there appears a
     detuning,
     $ \delta = \omega_0 - \omega_d $,
     where $ \hbar \omega_0 $ is the energy difference between the two internal
     states, and  $ \omega_d = \omega_L - \omega_S $ is the frequency of the
     Raman drive.
         Following Ref. [6]
 the boson field operators can be represent 
by an expansion in the Wannier basis,
which is a sum
 over 
  the lattice $ i $,   
         $ \Psi_a({\bf x}) = \sum_{ i } a_i w_0({\bf x} - {\bf x}_i),     
 \Psi_b({\bf x}) = \sum_{ i } b_i w_n({\bf x} - {\bf x}_i) , $ 
      where
 the operators $ a_i $ and $ b_i $ are the operators on lattice site $ i $;
    the annihilation and creation 
    operators
    $ a_i $ and $ a_i^{+} $ ( $ b_i $ and $ b_i^{+} $ ) obey
      the canonical commutation relations $ [ a_i, a_j^{+} ] = \delta_{ij}
   ( [ b_i, b_j^{+} ] = \delta_{ij} ) $.
  
 As a result, 
  in the tight - binding approximation  a two - band  Bose - Hubbard
  Hamiltonian of a strongly resonant driven BEC is 
     $$
     H_{BH} = -t_{aa} \sum_{<i,j>} a_i^{\dagger} a_j \varphi_a^2 -
    t_{bb} \sum_{<i,j>} b_i^{\dagger} b_j \varphi_b^2
     $$
     $$
    +\frac{U_{aa}}{2} \sum_i a_i^{\dagger} a_i ( a_i^{\dagger} a_i -
     1 ) \varphi_a^4+
    \frac{U_{bb}}{2} \sum_i b_i^{\dagger} b_i ( b_i^{\dagger} b_i -
    1 ) \varphi_b^4
  + U_{ab} \sum_i a_i^{\dagger} a_i b_i^{\dagger}
   b_i \varphi_a^2(t) \varphi_b^2+ 
     $$
     $$
 - \mu ( \sum_i a_i^{\dagger} a_i \varphi_a^2 + \sum_i b_i^{\dagger} b_i
  \varphi_b^2)
 + \delta \sum_i b_i^{\dagger} b_i \varphi_b^2,         \eqno ( 3 )
     $$
 where the sum in the two terms runs over all nearest neighbors denoted by
 $ < i,j > $. 
 The term involving the chemical potential $ \mu $ is added as the  
 calculation
 is in the grand - canonical ensemble at temperature $ T = 0 $.
    The first and second sum measure the boson kinetic energy [10],
     and the parameter
   $ t_{ll} $ is the hopping matrix element for channel $ l $
    between neighboring lattice sites
   $ i, j $;
   the parameters $ U_{aa}, U_{bb}, U_{ab} $ characterize the strength
 of the on - site interaction of two atoms on the lattice site $ i $.
 These parameters are defined as
   $$
   t_{aa} = \int d^{3}x w_0^{\ast}({\bf x} - {\bf x}_i) [ -\frac{ \hbar^2}{ 2 m}
   \nabla^2 + V_0({\bf x}) ] w_0({\bf x} - {\bf x}_j),
   $$
   $$
   t_{bb} = \int d^{3}x w_n^{\ast}({\bf x} - {\bf x}_j) [-\frac{\hbar^2}{ 2 m}
   \nabla^2 + V_0({\bf x}) ] w_n({\bf x} - {\bf x}_j), 
   $$
   $$
   U_{aa} = U_a^s \int d^{3}x | w_0({\bf x})|^4,
   U_{bb} = U_b^s \int d^{3}x | w_n({\bf x})|^4,
   U_{ab} = U_{ab}^s \int d^{3}x | w_0({\bf x})|^2 | w_n({\bf x})|^2.     \eqno (4) 
   $$ 
  The parameters $ t_{00} $ and $ U_{00} $ of the one - band Bose - Hubbard
  model
  depend on the potential depth $ V_o $, and they
   were calculated analytically in Ref.[11]
   in the tight - binding limit
   in terms of the microscopic parameters of the
  atoms in the optical lattice
   using the approximation [11]
 the Wannier states  by the harmonic oscillator wave function  
 $ w_n({\bf x} - {\bf x}_i) $ of number $ n $.

     In order to
   describe the zero - temperature quantum phase transition from the superfluid to 
   the
 Mott insulator phase analytically, it is neccesary to make some appropriate
 mean - field approximation in using the Hamiltonian ( Eq.(3)). A mean - field
 theory, developed in [11], constructs a consistent
 mean - field theory by substituting
 $
 a_i^{\dagger} a_j = \langle a_i^{\dagger} \rangle a_j + 
 a_i^{\dagger} \langle a_j \rangle - \langle a_i^{\dagger} \rangle
 \langle a_j \rangle,               
 $
 where
 $ \langle a_i^{\dagger} \rangle = \langle a_i \rangle $ is the expectation 
 value
 on site $ i $.
 Following this theory [11], which exploited the usual Landau procedure for a
 second - order phase transition, we obtain the critical value of the ratio
 of the on - site interaction and hopping parameters
  $$
  (\frac{U_{ll}}{t_{ll}})^{(cr)} \varphi_l^2 = 2 d ( 2 g + 1 + \sqrt{( 2g + 
  1)^2 - 1}),        \eqno (5)
  $$ 
 where the number of nearest neighbors $ 2 d ,  d $ is the number of 
 dimensions, 
 $ g $ is the integer number of particles at each site $  i $.
 In these two equations the functions $ \varphi_l $ coupled by the normalization
 condition. For a single channel ( one - component
 model ) $ \varphi_0 \equiv 1 $ and these equations
 are reduced to the well - known result for the critical value
 [11].
 Note that the chemical potential $ \mu_c $ at the critical point is calculated
 also by the cross - interaction $ U_{ab} $ between the channels.
 
   Let us first consider the time - dependent
   regime of the Rabi oscilations 
   for zero detuning, $ \delta = 0 $, in this case we have the time - dependent
   functions $ \varphi_l(t)$.
 In this regime the Eqs.(5) are valid if the Rabi oscillations period is
 larger than the nonequilibrium dynamical times (the characteristic
 time scales are
  the time of establishing of the SF state, the restoration time
    of the phase coherence or the dephasing time).
 The nonequilibrium dynamic time for trapped bosonic atoms in an optical
 lattice potential was considered in Ref. [12] and was shown that the
 characteristic time of the dynamical restoration of the phase coherence 
 in the domain of a quantum phase transition is the integer
 of the order
  of 10 multiplied
 by a Josephson time $ h/t $. 
    The
   initial phase of the BEC is MI phase for
   $ V_0 \succ V_0^{cr} $,  
   where $ V_0^{cr} $ is the optical potential depth at the
    SF - MI transition without driving.
       The functions
 $ \varphi_l(t)$ determine the population $ \rho_l $ of the states $ l $, which
   oscillate
     as $ \cos^2 \Omega_R t $ for channel $ a $
    and
     $ \sin^2 \Omega_R t $ for channel $ b $. 
  Accordingly, during the period $ 2 \pi/\Omega_R $ of the Rabi oscillations  
    the transition from MI to SF phases and the reverse transition occur. 
  Really, for the channel $ a $ at the point $ \Theta_1 $ there is the MI - SF
  transition, at $ (\pi - \Theta_1) $ - the opposite SF - MI transition.
    For times of the Rabi oscillations period from
  $ t = 0 $ until $ t_1 =\Theta_1/\Omega_R $ 
  and from $ (\pi - \Theta_1) $ until $ \pi $
  the channel $ a $ is in the MI phase,
   from $ \Theta_1 $ until $ (\pi - \Theta_1) $ there is the SF phase.
   Correspondingly,
   for the channel $ b $ with
  the opposite time - dependence of population
  at $ \Theta_2 $ there are the SF - MI transition, at $ (\pi - \Theta_2) $
  the MI - SF transition. In the channel $ b $,
   until $ t_2 = \Theta_2/\Omega_R $ and from $ (\pi - \Theta_2) $ until $ \pi $
  is in the SF phase, from $ \Theta_2 $ until
   $ (\pi - \Theta_2) $, there is the MI phase.
    Thus, both channels have the superfluid - Mott insulator transition twice in
    each half - Rabi period.
        The frequency of this oscillating superfluidity is determined by
      the driving strength.
       The time - averaged population
   $ N_{dr} $ of SF phase of a driven
    BEC, which without driving is in the MI phase,  is
   $$
   \frac{ N_{dr}( V_0 )}{ N_0} = \frac{ \pi + 2 [ \Theta_2( V_0) -
   \Theta_1( V_0) ] -
    [ \sin 2\Theta_1( V_0) + \sin 2\Theta_2( V_0) ]}{ 2 \pi},     \eqno (6)
   $$    
  where $ N_0 $ is the concentration of a nondriven 
  BEC in the SF phase for $ V_0 \prec V_0^{cr} $.
  Here
  $$
  \Theta_1( V_0) = \arccos \sqrt \frac { \Delta U_{00}( V_0)}
  { \Delta t_{00}( V_0)},
  \Theta_2( V_0) = \arcsin \sqrt 
  {\frac { \Delta U_{00}( V_0)}{ \Delta t_{00}( V_0)}
   \frac{ \Delta U_{00}^{nn}}{ \Delta t_{00}^{nn}}},      \eqno (7)
  $$
  where $ \Delta U_{00}(V_0) \equiv \frac{U_{00}(V_0^{cr})}{U_{00}(V_0)} ,
  \Delta t_{00}(V_0) \equiv \frac{t_{00}(V_0^{cr})}{t_{00}(V_0)}; $
     the ratios,
    which are calculated in 
     Eqs.(10) below for $ n = 1 $, are
 $ \Delta U_{00}^{nn} \equiv \frac{ U_{nn}}{ U_{00}},
  \Delta t_{00}^{nn} \equiv \frac{ t_{nn}}{ t_{00}}. $
  If $ \frac { \Delta U_{00}( V_0)}{ \Delta t_{00}( V_0)}
   \frac{ \Delta U_{00}^{nn}}{ \Delta t_{00}^{nn}} \geq 1 $ then
  $ \Theta_2( V_0) = \pi/2 $.

   In the steady state driving regime
    every atom is in
   the stationary 
   superposition of two states, which
   populations are determined by the Rabi frequency and the
   detuning from resonance. 
  We have
   two critical values
  of the transition parameters of the driven system 
  $ (U_{ll}/t_{ll})_{dr}^{(cr)} $,
  \{ l = { a, b } \}.
  The dependence  $ U_{ll}( V_0), t_{ll}( V_0) $ for every channel determines
   two
    optical potential depths
  $ V_{0(dr)}^{(a) {cr}} $
  and $ V_{0(dr)}^{(b) {cr}} $ at a phase transition, correspondingly.
   The critical value of parameter of a phase
  transition 
  for the driven system
   $ (U_{ll}/t_{ll})_{dr}^{(cr)} $ is always
   higher than without driving $ (U_{ll}/t_{ll})_0^{(cr)} $.
    For the channel $ a $
   $$
   \frac{(U_{00}/t_{00})_0^{(cr)}}{(U_{00}/t_{00})_{dr}^{(cr)}} = 1 -
   \frac{1}{2}[ 1 + (\frac{\delta}{\Omega_R})^2]^{-1}.         \eqno (8)
   $$ 
    Therefore, there are an interval of the MI phase, in which the driving
  switches  the MI phase to the SF phase. In this band  
  from $ (U_{00}/t_{00})_0^{(cr)} $ to $ 2 (U_{00}/t_{00})_0^{(cr)} $,
    by varying the 
   parameter $ \delta/\Omega_R $ of the external driving, we can tune the  
  critical value to a given value. With increasing the depth of the optical potential, the on - site interaction
  increases and the tunneling matrix element is reduced.
  Accodingly, with increasing the critical value of the transition parameter,
  the optical potential depth $ V_{0 (dr)}^{(0)cr} $ at a phase transition
 always increases.    
  
      For the channel $ b $ the critical value is always more than for the
      channel $ a $
     $$
     \frac{(U_{nn}/t_{nn})_{dr}^{(cr)}}{(U_{00}/t_{00})_{dr}^{(cr)}} =
     2 [ 1 + (\frac{\delta}{\Omega_R})^2] -1.          \eqno (9)
     $$
     We obtain for $ n = 1 $
  and in the tight - binding limit,
  $ \lambda \gg \beta $, the ratio
   $$
   \frac{ U_{11} }{ U_{00}} \simeq \frac{3}{8} H_{6}(0),
   \frac{t_{11}}{t_{00}} \simeq (\frac{\lambda}{ 2\beta})^2. \eqno (10) 
   $$ 
  Here $ \beta = \sqrt{\frac{\hbar}{ 2 m \nu_j}} $
 is the size of the ground - state oscillator 
 wave function determined by the atomic mass $ m $ and the oscillation frequency
 in the wells ( see, for example [ 6 ] ) $ \nu_j =\sqrt{ 4 E_R V_{jo} }/{\hbar} $,
  $ E_R $ is the recoil energy, 
  $ H_6(x) $ is Hermite polynomial. In general, the ratio 
  $ U_{nn}/U_{00} $ contains sum over the Hermite
   polynomials $ H_n(0) $   
     till $ 4n + 2 $, and 
    $ t_{nn}/t_{00} \simeq (\lambda/{2 \beta})^{2n} $,
 For all $ n $ there are the inequalities 
   $ t_{nn} \gg t_{00} $ and
   $ U_{nn} \gg U_{00} $,
   this increase $ t_{nn} $ and $ U_{nn} $ is due to 
 oscillations of the wave function for $ n \neq 0 $.
     For example,
 $ t_{11} / t_{00} \sim 64 $ and $ U_{11} / U_{00} \sim 45 $ for sodium [13]
 where $ E_R/{\hbar} \sim 2 \pi\times 10^4 Hz $ for a red detuned laser with
 $ \lambda \simeq 10^3 nm $ and $ V_0 \simeq 16 E_R $.
 Although, the on - site interaction on the higher channel $ U_{nn} $ increases
 with the number $ n $, the tunneling matrix element $ t_{nn} $
  increases faster, and
     we can say that the higher channel $ n $ is " more superfluid".
      The ratio $ U_{nn}/t_{nn} $ increases also with increasing the optical
   potential depth $ V_0 $, and because there are the inequalities 
   $ U_{nn}/t_{nn} \prec U_{00}/t_{00} $
    for the given
   $ V_0 $, then Eq.(9) yields a relationship 
   $ V_{0(dr)}^{(b)cr} \succ V_{0(dr)}^{(0)cr} $.

      In summary, the resonant strong driving by a spatially uniform field
   modify the SF - MI transition in a BEC in an optical lattice by the
   mechanism of the turning on of the second channel. Therefore, there are
   the modifications of the main channel     
     and the second channel. As a result, the transition
     on each channel
      is controlled by the external
     field. In the time - dependent driving regime the initial 
     MI phase switches over to the SF phase and the SF phase
     to the MI phase during
      the Rabi oscillations 
     period.
     The frequency of the oscillating superfluidity
     is determined by the
     driving strength.
      In the steady state driving regime, the critical values of the ratio
     of the on - site interaction and the atomic tunneling rate
     depend on
       the external field parameters ( the strength and the detuning
     from resonance). Two critical values of the driven system 
     and,
     accordingly,
       two optical potential depths at the phase transition 
      are always more than without driving.
          The proposed mechanism of manipulating and controlling Bose - Einstein
      condensates makes it possible to
       tune 
     the critical value 
     of parameters
     of the quantum phase transition to a given value.

           I thank 
           H. E. Stanley for encouragement.

\bibliographystyle{prsty}

\begin{thebibliography}{0}

\bibitem{1} M. H. Anderson {\it et al.}, Science {\bf 269}, 198 (1995). 

\bibitem{2} C. C. Bradley {\it et al.}, Phys. Rev. Lett. {\bf 75}, 1687 (1995).

\bibitem{3} K. B. Davis {\it et al.},
    Phys. Rev. Lett. {\bf 75}, 3969 (1995). 

\bibitem{4} C. Orzel {\it et al.}, Science {\bf 291}, 2386 (2001).

\bibitem{5} M. Greiner {\it et al.}, 
 Nature ( London ) {\bf 415}, 39 (2002).  


\bibitem {6} D. Jaksch {\it et al.}, Phys. Rev. Lett. {\bf 81}, 3108 (1998).  


\bibitem{7} J. Williams {\it et al.}, Phys. Rev. A {\bf 59}, R31 (1999).

\bibitem{8} G. M. Genkin, Phys. Rev. A {\bf 63}, 025602 (2001).

\bibitem{9} S. Kohler and F. Sols, Phys. Rev. Lett. {\bf 89},
060403 ( 2002).


\bibitem{10} Note that the terms 
         $ t_{0n} $ and $ t_{n0} $ are absent because the hopping
  matrix element $ t_{0n} $
  from the initial lowest vibrational state  to the
  ultimate state with $ n \neq 0 $ is identically equal
   to zero for $ T = 0 $. The
  element $ t_{n0} $ is much smaller than the element $ t_{nn} $ since the
  tunneling with emission the excitation
   contains the additional small parameter.

\bibitem{11} D. van Oosten, P. van der Straten, and H. T. C. Stoof,
Phys. Rev. A {\bf 63}, 053601 ( 2001).

\bibitem{12} A. Polkovnikov, S. Sachdev, and S. M. Girvin, Phys. Rev.
 A {\bf 66}, 053607 ( 2001).
                    
 \bibitem{13} D. M. Stamper-Kurn {\it et al.}, Phys. Rev. Lett. {\bf 80},
         2027 ( 1998).
                     

\end{thebibliography}

\newpage                        




\end{document}